\documentclass[conference]{IEEEtran}
\IEEEoverridecommandlockouts
\usepackage[utf8]{inputenc}
\usepackage{cite}
\usepackage{floatrow}
\usepackage{amsmath,amssymb,amsfonts,stfloats}
\usepackage{tabularx}
\usepackage{algorithmic}
\usepackage{graphicx}
\usepackage{textcomp}
\usepackage{floatrow}
\usepackage{xcolor}
\usepackage{booktabs}

\usepackage{multicol}
\usepackage{acronym}
\usepackage{url}
\usepackage{mathtools}
\usepackage[font=small]{caption}
\usepackage[caption=false,font=footnotesize]{subfig}

\def\BibTeX{{\rm B\kern-.05em{\sc i\kern-.025em b}\kern-.08em
    T\kern-.1667em\lower.7ex\hbox{E}\kern-.125emX}}

\newcommand{\linebreakand}{%
  \end{@IEEEauthorhalign}
  \hfill\mbox{}\par
  \mbox{}\hfill\begin{@IEEEauthorhalign}
}

\title{Waveform Design for OFDM-based ISAC Systems Under Resource Occupancy Constraint}

\author{\IEEEauthorblockN{Silvia Mura\IEEEauthorrefmark{1}, Dario Tagliaferri\IEEEauthorrefmark{1}, Marouan Mizmizi\IEEEauthorrefmark{1}, Umberto Spagnolini\IEEEauthorrefmark{1}, and Athina Petropulu\IEEEauthorrefmark{2}}
	\IEEEauthorblockA{\IEEEauthorrefmark{1}Politecnico di Milano, Milan, Italy}\IEEEauthorblockA{\IEEEauthorrefmark{2} Rutgers University., Piscataway, USA} 
	E-mail (corresponding):\,silvia.mura@polimi.it}

\begin{document}

\maketitle

\begin{abstract}
Integrated Sensing and Communication (ISAC) is one of the key pillars envisioned for 6G wireless systems. ISAC systems combine communication and sensing functionalities over a single waveform, with full resource sharing. In particular, waveform design for legacy Orthogonal Frequency Division Multiplexing (OFDM) systems consists of a suitable time-frequency resource allocation policy balancing between communication and sensing performance. Over time and/or frequency, having unused resources leads to an ambiguity function with high sidelobes that significantly affect the performance of ISAC for OFDM waveforms. This paper proposes an OFDM-based ISAC waveform design that takes into account communication and resource occupancy constraints. The proposed method minimizes the Cramér-Rao Bound (CRB) on delay and Doppler estimation for two closely spaced targets. Moreover, the paper addresses the under-sampling issue by interpolating the estimated sensing channel based on matrix completion via Schatten $p$-norm approximation. Numerical results show that the proposed waveform outperforms the state-of-the-art methods.
\end{abstract}


\vspace{-0.1cm}

\section{Introduction}

In the pursuit of advancing wireless communication technologies, the integration of sensing capabilities has emerged as a paradigm shift, offering unprecedented opportunities for the development of intelligent and context-aware communication systems~\cite{Saad2020AVO}. The cohesive integration of communication and sensing, recognized as Integrated Sensing and Communication (ISAC), significantly influences the architecture of forthcoming 6G networks, holding the potential to enhance performance, energy efficiency, and multiple novel applications \cite{wild2021JCS6G}.

%

The development of a unified ISAC waveform for both data transfer and sensing has been identified as a substantial research challenge in recent years~\cite{b1}. Adapting legacy Orthogonal Frequency Division Multiplexing (OFDM) for this purpose aims to provide a simple yet effective ISAC implementation. The work in~\cite{b2} proposes partitioning OFDM subcarriers for radar and communication, introducing a trade-off in the distinct allocation of subcarriers. However, introducing radar-dedicated subcarriers for sensing purposes requires an energy expenditure that can be avoided with a suitable optimized resource allocation for communication \textit{and} sensing. Conversely,~\cite{xu2023bandwidth} introduces an OFDM waveform optimizing bandwidth for communication and sensing, enabling antennas to transmit across multiple subcarriers categorized into shared and private groups.

Existing studies employ various metrics to formulate the ISAC resource allocation problem in OFDM-based systems. The paper in~\cite{b4} introduces an OFDM chirp waveform that incorporates embedded phase-modulated communication information tailored for delay-Doppler radar systems. 
The authors of~\cite{b5}, instead, present an optimal allocation of power within the frequency-time (FT) resources, incorporating the ambiguity function as a constraint in the Cramér-Rao bound (CRB) minimization on delay and Doppler estimation of one target. The study includes a trade-off analysis between sensing performance and communication capabilities through CRB and capacity optimization, specifically applicable to scenarios involving full bandwidth occupancy and a single target.

The previously mentioned works do not address the performance of the ISAC waveform in scenarios where the available FT resources are not fully utilized (i.e., resource occupancy factor $<100\%$). It is crucial to emphasize that, in practice, FT resources rarely reach their full utilization (only during moments of extreme congestion). Herein, we introduce a maximum resource occupancy factor to reflect operational reality, providing a more accurate perspective on the performance of the ISAC signal in the everyday usage scenarios of wireless networks (e.g., refer to the 3GPP standard~\cite{b6}). Lower occupancy factors lead to high sidelobes in the ambiguity function. This negatively impacts the sensing performance, especially when detecting two or more close targets \cite{b5}.
%
%
%
The work~\cite{b8} proposes a waveform design method based on minimizing delay and Doppler CRBs by filling empty subcarriers with sensing samples at the cost of increased energy consumption. 
Differently, the authors of~\cite{b7} introduce a linear interpolation across OFDM symbols to enhance target detection by addressing unused subcarriers. However, this approach fails when the allocated FT resources are highly sparse.


\textit{Contribution}: This paper introduces a novel waveform design within OFDM-based ISAC systems, considering the constrained utilization of FT resources. The primary objective is to minimize the CRB associated with delay and Doppler estimation for two coupled targets. The method operates under constraints imposed by spectral efficiency and the resource occupancy factor. To address the resulting high sidelobes, a matrix completion-based interpolation technique is proposed for the estimated sensing channel in the FT domain. Comparative analysis against the conventional approach, employing random scheduling in conjunction with the linear interpolation in~\cite{b7}, reveals the superior performance of the proposed waveform and interpolation technique. Specifically, our suggested method achieves the CRB for delay and Doppler estimation for high signal-to-noise ratios (SNRs) with limited bandwidth, while the conventional one encounters difficulties due to the under-sampling.

The following notation is adopted in the paper: bold upper- and lower-case letters describe matrices and column vectors. The $ij$-th entry of $\mathbf{A}$ is $[\mathbf{A}]_{ij}$. Matrix transposition, conjugate transposition and $L$-norm are indicated respectively as $\mathbf{A}^T$, $\mathbf{A}^H$ and $\|\mathbf{A}\|_L$. The element-wise product between two matrices is denoted by $\odot$. $\mathrm{diag}(\mathbf{A})$ denotes the extraction of the diagonal of $\mathbf{A}$. $\mathrm{vec}(\mathbf{A})$ denotes the vectorization by columns of $\mathbf{A}$ and $\mathrm{vec}^{-1}(\cdot)$ denote the inverse operation. $\mathbf{1}_N$ is a columns vector of $N$ entries equal to one. With  $\mathbf{a}\sim\mathcal{CN}(\boldsymbol{\mu},\mathbf{C})$ we denote a multi-variate circularly complex Gaussian random variable with mean $\boldsymbol{\mu}$ and covariance $\mathbf{C}$. $\mathbb{E}[\cdot]$ is the expectation operator, while $\mathbb{R}$, $\mathbb{C}$ and $\mathbb{B}$ stand for the set of real, complex, and Boolean numbers, respectively. $\delta_{n}$ is the Kronecker delta.

The remainder of the paper is organized as follows: Section \ref{sec:system_model} outlines the system model, Section \ref{sec:waveform_design} presents the proposed waveform design, and Section \ref{sect:ch_interp} discusses the proposed channel estimation and interpolation. Numerical results are presented in Section \ref{sect:results}, while Section \ref{sect:conclusion} concludes the paper.

\section{System Model}\label{sec:system_model}

\begin{figure}[b!]
    \centering
    \includegraphics[width=0.75\columnwidth]{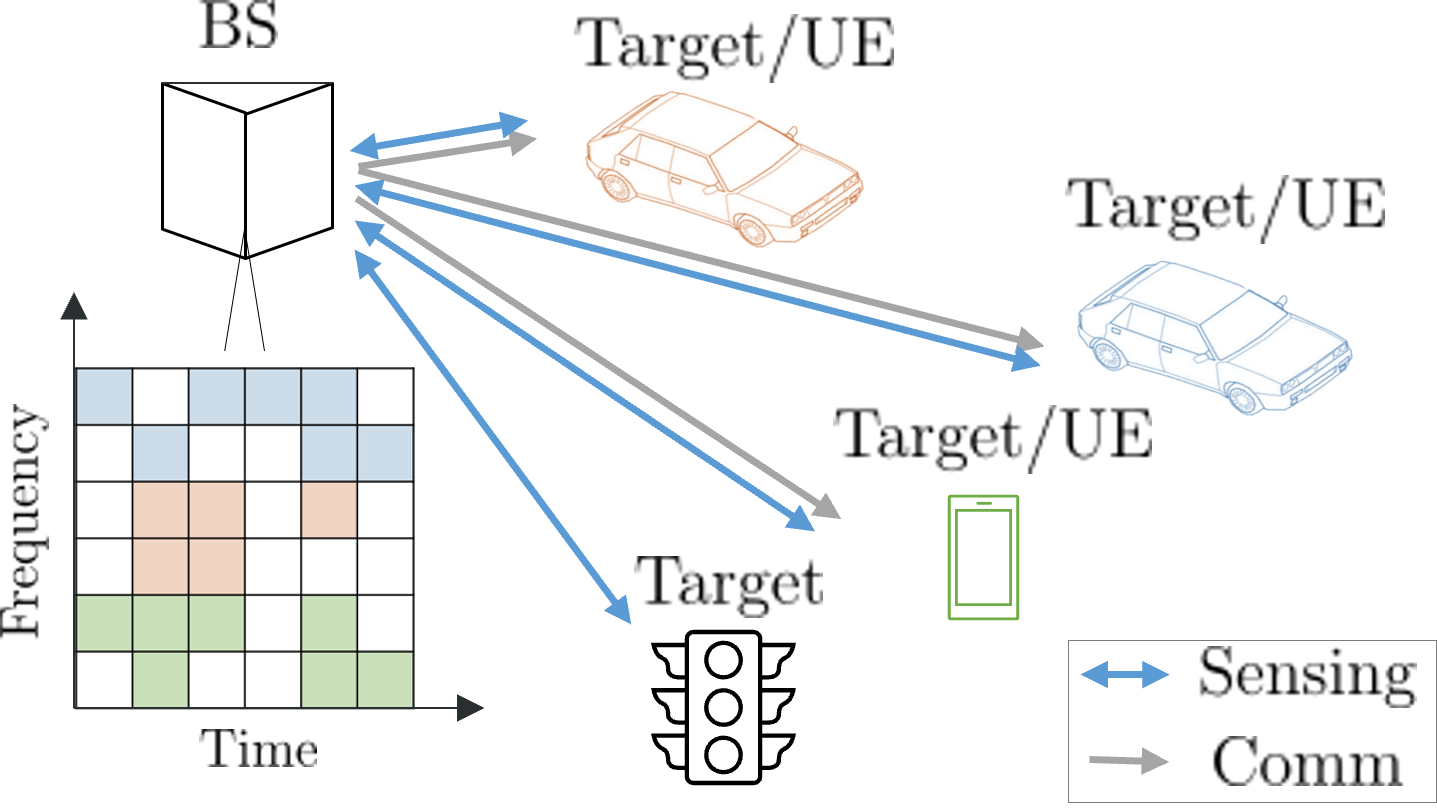}
     \caption{ISAC waveform design scenario with multiple targets/UEs.}\label{fig:scenario}
\end{figure}

We consider the ISAC system in Fig. \ref{fig:scenario}, where the base station (BS) is tasked with providing service to a set of $K$ users (UEs) while simultaneously estimating their delay and Doppler shifts.  We assume, for simplicity, that the UEs to be served are the sole targets of interest within the operating environment. In this work, we focus on FT ISAC waveform design, therefore any spatial precoding/decoding at the Tx/Rx antennas is implicit. The ISAC BS uses an OFDM waveform in which the available time and frequency resources for downlink communication and sensing tasks are organized in $M$ subcarriers and $N$ time slots, respectively spaced by $\Delta f$ (total bandwidth $B = M \Delta f$) and  $T = 1/\Delta f$ (ISAC burst duration $NT$).  
The $mn$-th element of the $M\times N$ Tx signal matrix over the discrete frequency-time domain is
\begin{equation}
    [\mathbf{X}]_{mn} = [\mathbf{P} \odot \mathbf{S}]_{mn} = P_{mn} s_{mn}
\end{equation}
where $[\mathbf{P}]_{mn}=P_{mn}\in \mathbb{R}_+$ is the square power of the allocated power to the $mn$-th communication symbol $[\mathbf{S}]_{mn}=s_{mn}\in \mathbb{C}$, the latter drawn from an arbitrary constellation with unitary power ($\mathbb{E}[s_{mn}]=0, \mathbb{E}[|s_{mn}|^2]= \delta_{m-m'}\delta_{n-n'}$).

\subsection{Rx signal at the BS}

The Rx signal matrix $\mathbf{R} \in \mathbb{C}^{M \times N}$ at the BS in the time-frequency domain is:
%
%
\begin{equation}\label{eq:RxsignalBS}
    \mathbf{R} = \mathbf{X} \odot \mathbf{H}_s + \mathbf{W},
\end{equation}
where $\mathbf{H}_s \in \mathbb{C}^{M  \times N}$ is the sensing channel collecting the echoes from the $K$ UEs/targets and $\mathbf{W} \in \mathbb{C}^{M \times N}$ collects the noise samples in frequency-time domain, such that $[\mathbf{W}]_{mn}=w_{mn} \sim\mathcal{CN}(0, \sigma^2_w\delta_{m-m'}\delta_{n-n'})$, which is uncorrelated across time and frequency. The sensing channel $\mathbf{H}_s \in \mathbb{C}^{M  \times N}$ at the $mn$th resource bin is modelled as: 
\begin{align}\label{eq:sensing_channel_FT}
 [\mathbf{H}_s]_{mn} = \sum_{k=1}^{K} \beta_{k} \, e^{j 2 \pi (\nu_{k} n T-\tau_{k} m \Delta f)},   
\end{align} 
where \textit{(i)} $\beta_{k} \sim \mathcal{CN}(0,\Omega^{(k)}_{\beta})$ represents the complex scattering amplitude of the $k$-th UE/target, whose power $\Omega^{(k)}_{\beta} \propto f_0^{-2} R_k^{-4}$ is function of the carrier frequency $f_0$ as well as of the distance between the BS and UE/target $R_k$, including the target's reflectivity, \textit{(ii)} $\tau_{k}=2 R_k/c$ corresponds to the propagation delay associated with the $k$th UE/target, \textit{(iii)} $\nu_{k}= 2 f_0 V_k/c$ represents the Doppler shift corresponding to radial velocity $V_k$. Models \eqref{eq:RxsignalBS} and \eqref{eq:sensing_channel_FT} are valid under the assumption that the maximum delay $\tau_{max} = \mathrm{max}_k\left(\tau_k\right)$ is less than the duration of the employed cyclic prefix $T_{cp}$, thus $\tau_{max}\leq T_{cp}$, to enable unambiguous range estimation~\cite{Tagliaferri_TWC}.

\subsection{Rx signal at the UE}

The model for the time-frequency Rx signal at the $k$-th UE on the $mn$-th frequency-time resource bin is
%
%
\begin{align}
      \mathbf{Y}_k = \mathbf{X} \odot \mathbf{H}_{k} + \mathbf{Z},  
\end{align}
with the communication channel 
\begin{align}
 [\mathbf{H}_k]_{mn} =\sum_{q=1}^{Q} \alpha_{q}^{(k)} \,e^{j 2 \pi( {\nu}_{q}^{(k)} nT - m\Delta f {\tau}_{q}^{(k)})},   
\end{align} 
%
%
where $Q$ refers to the number of paths, assumed to be the same for all the UEs for the sake of simplicity, $\alpha_{q}^{(k)} \sim \mathcal{CN}(0,\Omega^{(k)}_{q})$ is the complex amplitude of the $q$-th path, $k$-th UE, ${\tau}_{q}^{(k)}$ and ${\nu}_{q}^{(k)}$ denote the delay and Doppler shift of the $q$-th path, $k$-th UE, respectively. However, unlike the Rx signal $\mathbf{R}$ at the BS, the communication channel does not preserve the true delay and Doppler shifts due to the time-frequency synchronization performed at the UE terminal. The additive noise is uncorrelated across UEs and across time and frequency, namely $z^{(k)}_{mn} \sim\mathcal{CN}(0, \sigma^2_z\delta_{m-m'}\delta_{n-n'}\delta_{k-\ell})$.

\section{Waveform Design}\label{sec:waveform_design}

The proposed ISAC waveform is  designed by selecting OFDM communication resources to minimize the weighted average of the delay and Doppler CRBs, while guaranteeing the quality of service (QoS) in terms of communication spectral efficiency (SE). We herein focus on the two targets/UEs case, i.e., $K = 2$, being representative of a situation in which the BS aims at serving two closely spaced UEs, while estimating their delay-Doppler signature. The CRB corresponding to the delay and Doppler estimates of the two targets is derived in the appendix \ref{app:CRB}.

Let us  indicate the  FT resources allocated to  the $k$-th UE by a boolean vector $\mathbf{a}_k \in \mathbb{B}^{L \times 1}$, where $L=MN$ is the total number of time-frequency resources, and
\begin{equation}\label{eq:allocation_vector}
    [\mathbf{a}_k]_{\ell} = \begin{cases} 1,\,\,\text{if the FT resource is chosen}, \\
    0,\,\,\text{otherwise}.
    \end{cases}
\end{equation}
Let the constant per-resource power be denoted by $\sigma^2$.  The overall waveform can be written as
\begin{equation}\label{eq:designed_waveform}
    \mathbf{X} = \sigma \mathbf{1}_M \mathbf{1}_N^T  \odot \mathbf{S} \odot \mathbf{A} ,
\end{equation} 
with $\mathbf{A} = \mathrm{vec}^{-1}\left(\sum_{k=1}^K \mathbf{a}_k\right)$ being the matrix of allocated resources over frequency and time.
The general optimization problem is formulated as follows:
\begin{subequations}
\begin{alignat}{2} 
&\underset{\mathbf{a}_k}{\mathrm{minimize}}  &\quad&  \epsilon_{\tau} \,\mathrm{tr}\left(\frac{\mathbf{C}_{\tau}(\mathbf{a}_k)}{\Delta \tau^2}\right) + \epsilon_{\nu}\mathrm{tr}\,\left(\frac{\mathbf{C}_{\nu}(\mathbf{a}_k)}{\Delta \nu^2}\right)\,\label{eq:optProb1}\\
&\mathrm{s.\,t.} &      & \frac{1}{L}\sum_{\ell=1}^L\log_2(1+{\gamma}([\mathbf{a}_k]_\ell)\geq  \overline{\eta},\,\,\, \forall k, \label{eq:prob1_constraint1}\\
&  &      & \sum_{k=1}^K [\mathbf{a}_k]_{\ell} \leq 1,\,\,\, \forall \ell, 
\label{eq:prob1_constraint2}\\
&  &      & \sum_{k=1}^K \mathbf{1}^T_{L}\mathbf{a}_k  \leq \mu L, \,\,\label{eq:prob1_constraint3}\\
&  &      & 1 \leq \ell \leq L, \nonumber\\,
&  &      & 1 \leq k \leq K \nonumber.
\end{alignat}
\end{subequations}
where $\mathbf{C}_{\tau}(\mathbf{a}_k) \in \mathbb{R}^{K\times K}$ and $\mathbf{C}_{\nu}(\mathbf{a}_k) \in \mathbb{R}^{K\times K}$ denote CRB matrices on the delay and Doppler estimation for the $K$ UEs/targets, function of the allocated resources as shown in Appendix \ref{app:CRB}. In the cost function, the CRBs are scaled by the maximum delay and Doppler resolution, respectively $\Delta \tau=1/B$ and $\Delta \nu=1/T$. This allows to have comparable quantities and dimensionless weights $\epsilon_\tau$ and $\epsilon_\nu$. 
In \eqref{eq:optProb1}, the objective is to minimize the sum of the CRBs, properly weighted by $\epsilon_{\tau}$ and $\epsilon_{\nu}$ respectively. Constraint \eqref{eq:prob1_constraint1} sets the QoS requirement, expressed with a SE threshold $\overline{\eta}$ in [bits/s/Hz] equal for all the UEs, where
\begin{align}
{\gamma}([\mathbf{a}_k]_\ell)= \frac{\sigma^2\,\,[\mathbf{a}_k]_\ell \,\,\|\boldsymbol{\alpha}_k\|_2^2}{\sigma_z^2},
\end{align}
is the communication SNR for each FT resource bin for the $k$-th UE, with $\boldsymbol{\alpha}_k$ =$[ \alpha_{1}^{(k)}, ...,  \alpha_{Q}^{(k)}]$.The SE constraint is linked to the quantity of selected resources for each UE, as the problem enforces a constant power allocation per resource. Consequently, when changing the distance between two targets, the one subject to higher path-loss is allocated a greater number of resources. Constraint \eqref{eq:prob1_constraint2} is enforced to avoid multi-user interference by assigning each time-frequency resource to a single UE. The total number of allocated resources is subject to the occupancy constraint dictated by \eqref{eq:prob1_constraint3}, where $\mu<1$ indicates the maximum fraction of resources that can be allocated.
 
The optimization problem \eqref{eq:optProb1} is non-convex in the objective function w.r.t. optimization variables, and it is, therefore, difficult to solve in the present form. However, the objective function can be relaxed according to \cite{b5}. In particular, the optimization problem resulting from the relaxation of the objective function can be classified as a mixed-integer conic programming problem (MICP), considering the binary nature of the model. These problems present inherent complexity, stemming from the incorporation of both continuous and discrete variables, supporting the application of the branch-and-cut (BnC) method. To mitigate the computational complexity linked with the BnC algorithm, \eqref{eq:optProb1} can be downscaled by considering frequency and time resources in grouped units referred to as subchannels and time slots, aligning with the framework established in the 3GPP standard \cite{b6} and complemented by appropriately adjusting the constraint in \eqref{eq:prob1_constraint1}. 

\begin{figure*} [t!]
    \centering
    \subfloat[][${\epsilon}_{\tau} = {\epsilon}_{\nu}= 0.5$]{ \includegraphics[width=0.25\columnwidth]{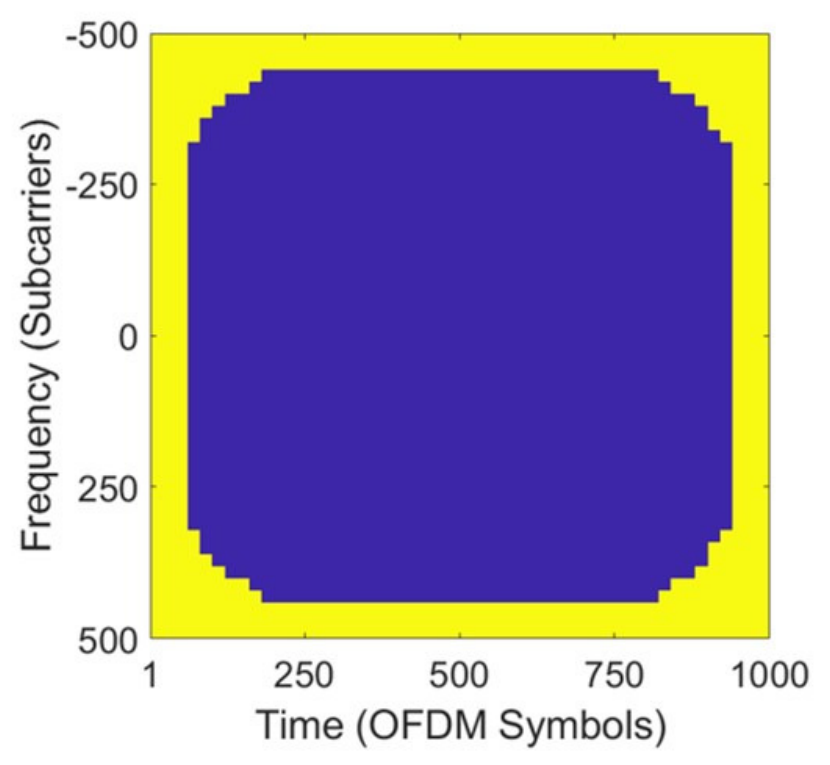}\label{fig:waveformb1}}
    \subfloat[ ][${\epsilon}_{\tau} = {\epsilon}_{\nu}= 0.5$]{\includegraphics[width=0.25\columnwidth]{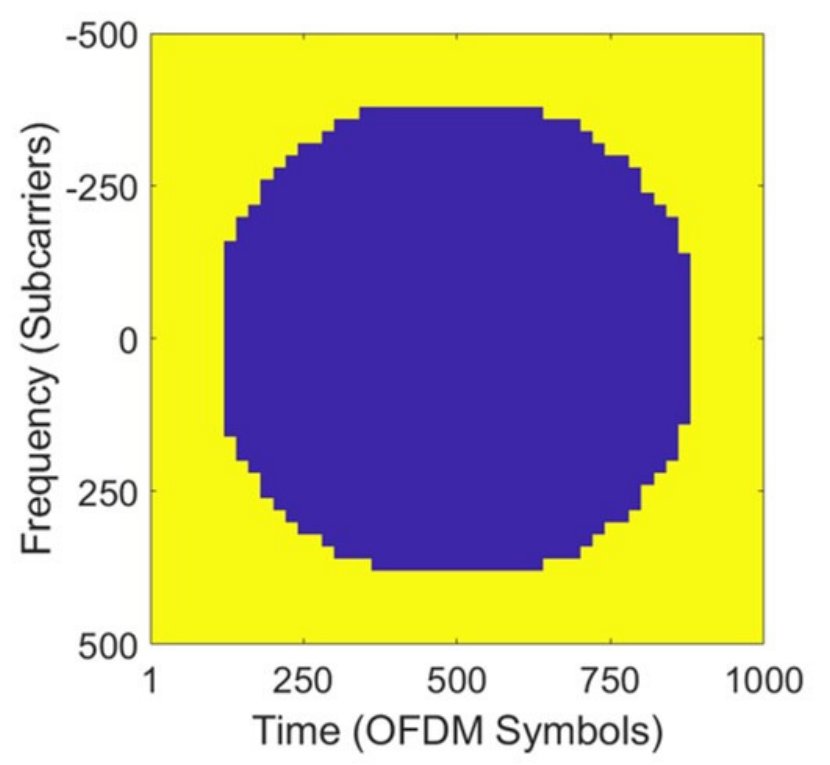}\label{fig:waveformb2}}
     \subfloat[][${\epsilon}_{\tau} = 0.25, {\epsilon}_{\nu} = 0.75$]{ \includegraphics[width=0.25\columnwidth]{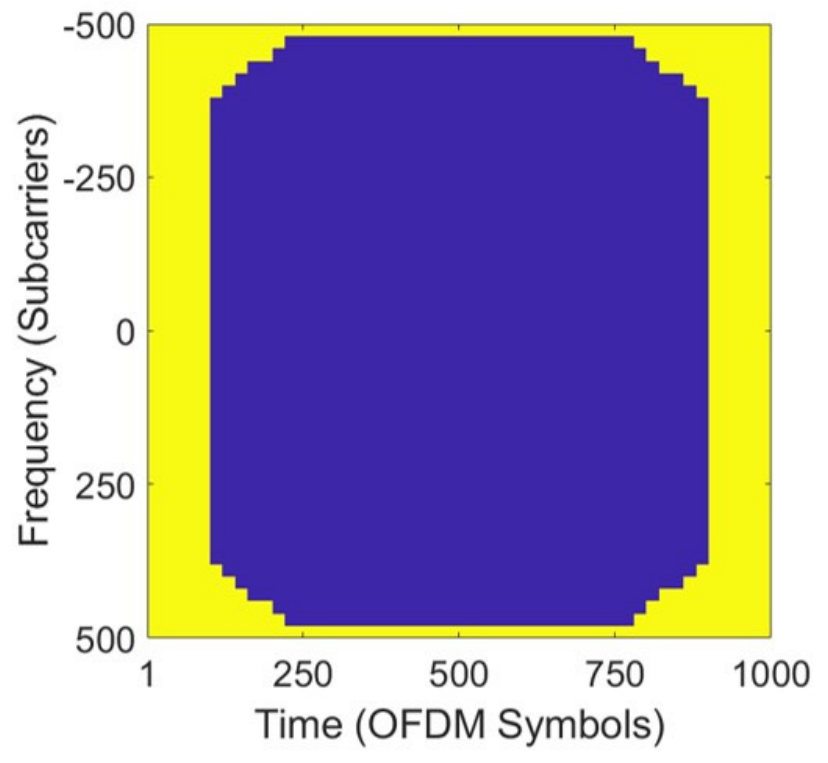}\label{fig:w1}}
    \subfloat[ ][${\epsilon}_{\tau} = 0, {\epsilon}_{\nu} = 1$]{\includegraphics[width=0.25\columnwidth]{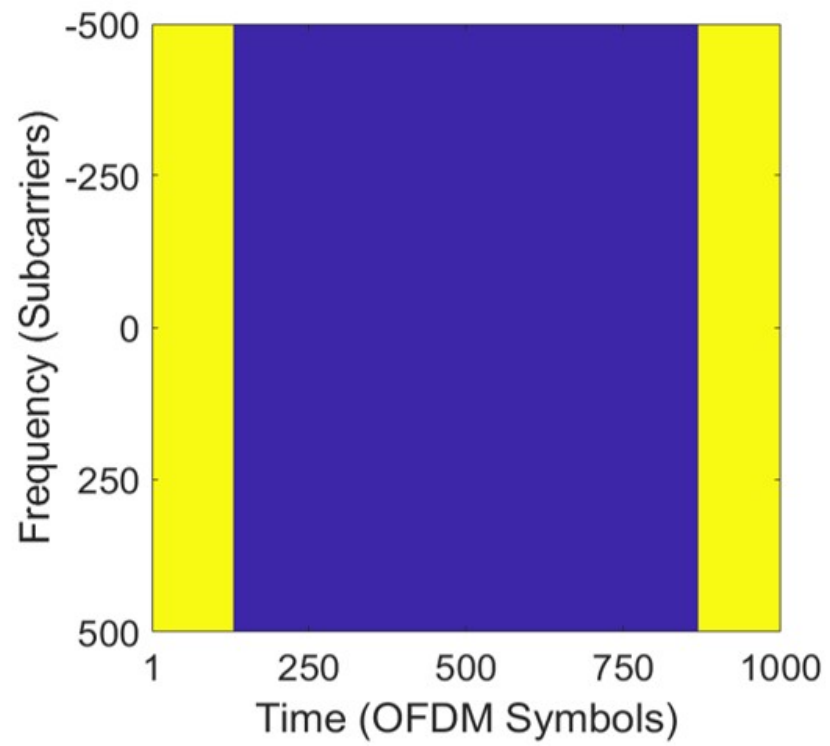}\label{fig:w2}}
    \caption{Optimized waveform for: (a) $\mu = 0.25$, ${\epsilon}_{\tau} = {\epsilon}_{\nu}= 0.5$ (b) $\mu = 0.5$, ${\epsilon}_{\tau} = {\epsilon}_{\nu}= 0.5$
    (c) $\mu = 0.25$,${\epsilon}_{\tau} = 0.25$, ${\epsilon}_{\nu} = 0.75$ (d) $\mu = 0.25$, ${\epsilon}_{\tau} = 0, {\epsilon}_{\nu} = 1$. Yellow denotes allocated resources, while blue denotes empty resource bins.}
    \label{fig:waveformb}
\end{figure*}

 As an example, Fig. \ref{fig:waveformb} shows the selected resources in the FT domain, showcasing the dependency on the resource occupancy factor $\mu$, for different weights $\epsilon_\tau$ and $\epsilon_\nu $. To minimize the CRB while satisfying communication constraints, the resource allocation gravitates towards the boundaries of the FT grid. Weights $\epsilon_{\tau}$ and $\epsilon_{\nu}$ affect delay and Doppler CRBs trade-off within the optimization process. Figure \ref{fig:waveformb1} illustrates the waveform achieved when both factors are balanced in the context of the CRB across the delay and Doppler dimensions. Conversely, Fig. \ref{fig:w1} portrays a scenario where greater emphasis is placed on minimizing the CRB of Doppler ($\epsilon_{\tau} = 0.25$, $\epsilon_{\nu} = 0.75$), while Fig. \ref{fig:w2} considers an extreme condition where the CRB on delay estimation is not taken into account ($\epsilon_{\tau} = 0$). This latter case reduces the optimization to minimizing the Doppler CRB, allocating all available resources along the frequency axis.

\section{Sensing Channel Interpolation and Parameters Estimation}\label{sect:ch_interp}

Delay and Doppler shift estimation follows from maximum likelihood (ML). By expressing the Rx signal \eqref{eq:RxsignalBS} in vector form, i.e., $\mathbf{r}=\mathrm{vec}(\mathbf{R})$, the ML formulates as:
\begin{equation} (\widehat{\boldsymbol{\tau}},\widehat{\boldsymbol{\nu}}) = \underset{\boldsymbol{\tau},\boldsymbol{\nu} }{\mathrm{argmin}} \left(\left\|\mathbf{r} - \mathbf{x}\odot \mathbf{h}_s(\boldsymbol{\tau},\boldsymbol{\nu} )\right\|^2_2\right),
\end{equation}
where $\boldsymbol{\tau} = [\tau_1,...,\tau_K]^T$ and $\boldsymbol{\nu} = [\nu_1,...,\nu_K]^T$ denote vectors of delays and Doppler shifts to be estimated. The ML estimation can be approached with an exhaustive search over the delay and Doppler shift domains, but the estimation in practical systems is approached in a suboptimal way, following three cascaded steps: \textit{(i)} estimation of the sensing channel $\mathbf{H}_s$ \textit{(ii)} 2D-DFT transform to map the estimated sensing channel $\mathbf{H}_s$ from FT domain to DD and \textit{(iii)} estimation of delays and Doppler shifts $(\boldsymbol{\tau},\boldsymbol{\nu})$. The estimation of the FT channel matrix over the allocated resources can be achieved through the least-squares approach, as follows:
\begin{align}\label{eq:estimated_channel}
    [\widehat{\mathbf{H}}_s]_{mn} = \begin{dcases}
        [\mathbf{R}]_{mn} \odot [\mathbf{X}]^{-1}_{mn} & \text{for $[\mathbf{A}]_{nm}=1$}\\
        0 &\text{for $[\mathbf{A}]_{nm}=0$}
    \end{dcases}
\end{align}
where $\mathbf{X}$ is the designed waveform according to \eqref{eq:designed_waveform}. Then the channel is mappend into DD domain through an DFT-IDFT pair as follows
\begin{equation}\label{eq:sensing_channel_DD}
   \widetilde{\mathbf{H}}_s = \mathbf{\Theta}_M \widehat{\mathbf{H}}_s \mathbf{\Theta}_N^H
\end{equation}
where $\mathbf{\Theta}_M\in\mathbb{C}^{M\times M}$ and $\mathbf{\Theta}_N\in\mathbb{C}^{N\times N}$ are DFT matrices such that $\|\mathbf{\Theta}_M\|_F=\sqrt{M}$ and $\|\mathbf{\Theta}_N\|_F=\sqrt{N}$. From the channel in DD domain, $(\boldsymbol{\tau},\boldsymbol{\nu})$ are estimated by multiple peak detection. Combination of steps \textit{(ii)} and \textit{(iii)} is substantially a 2D periodogram. However, when resource occupancy factor $\mu$ is significantly $<100\%$, the parameter estimation with the 2D periodogram does not work, especially for possibly coupled targets in the DD domain (e.g., UEs with similar velocity and range). Indeed, the estimated DD channel $\widetilde{\mathbf{H}}_s$ is the Fourier transform along frequency and time of a combination of power-scaled 2D sinusoids sampled over the allocated subcarriers $\mathbf{A}$. Therefore, multiple ghost peaks in the estimated channel $\widetilde{\mathbf{H}}_s$ arise due to the unused portions over the FT grid, possibly masking weakly scattering targets in the vicinity---in the DD domain--- of the comparably stronger one\footnote{This can also be devised by inspection of the ambiguity function of the Tx waveform $\mathbf{X}$.}. To obviate the latter issue, we propose an interpolation of the sensing channel in frequency and time, based on matrix completion. The matrix completion problem is formulated as follows
\begin{subequations}
\begin{alignat}{2} 
&\underset{{\mathbf{H}_s}}{\mathrm{minimize}}  &\quad& \mathrm{rank}({\mathbf{H}_s})\label{eq:optkp1Prob_interp}\\
&\mathrm{s.\,t.} &  &[{\mathbf{H}_s}]_{mn} =  [\widehat{\mathbf{H}}_s]_{mn} \,\,\, \text{for $[\mathbf{A}]_{nm}=1$}\label{eq:interpconstraint}.
\end{alignat}
\end{subequations}
Minimizing matrix rank under linear (affine) constraints is known to be NP-hard, but is often tackled using nuclear norm minimization, which is however a strong approximation of channel rank, thus not suitable for the typical resource occupancy factors considered in the paper (see e.g., \cite{b14} for further details). 
Conversely, we approach problem \eqref{eq:optkp1Prob_interp} by relaxing the objective function using Schatten $p$-norm. The latter is defined as 
\begin{equation}\label{eq:schatten}
\|\mathbf{H}_s\|_p^p = \left(\sum_{i}^{\mathrm{min}(M,N)} \Sigma_i^p\right)^{\frac{1}{p}}
\end{equation}
for $p \in (0,1]$, where $\Sigma_i$ is the $i$-th singular value of $\mathbf{H}_s$. Schatten $p$-norm offers a flexible trade-off between convexity and rank approximation in matrix optimization. When $p = 1$, the Schatten norm degenerates to the nuclear norm, and as $p$ approaches $0$, Schatten norm is a good rank approximator. This characteristic allows for balancing the complexity of optimization with recovery quality in matrix rank minimization problems. Therefore, the matrix completion problem can be rewritten as
\begin{subequations}
\begin{alignat}{2} 
&\underset{{\mathbf{H}_s}}{\mathrm{minimize}}  &\quad& \|\mathbf{H}_s\|_p^p\label{eq:optkp1Prob_interp1}\\
&\mathrm{s.\,t.} &  &[{\mathbf{H}_s}]_{mn} =  [\widehat{\mathbf{H}}_s]_{mn} \,\,\, \text{for $[\mathbf{A}]_{nm}=1$}\label{eq:interpconstraint1}.
\end{alignat}
\end{subequations}
Even though problem \eqref{eq:optkp1Prob_interp1} is still non-convex, it can be effectively addressed using numerically efficient algorithms like the iterated soft thresholding method proposed in \cite{b13}. Furthermore, the requisite condition for the unique recovery of any matrix $\mathbf{H}_s$ with a rank not exceeding $K$ via \eqref{eq:optkp1Prob_interp1} is established in \cite{b14}.

\section{Numerical Results}\label{sect:results}
This section validates the performance of the proposed ISAC waveform. For comparative analysis, our benchmarks comprise the random resource scheduling and the random scheduling with contiguous resources, both operating at the identical occupancy factor $\mu$ and utilizing the linear interpolation scheme in \cite{b7}. The standard random resource scheduling involves the random allocation of FT resources across the bandwidth, whereas the random contiguous resource scheduling allocates blocks of $N_b$ contiguous resources randomly within the bandwidth.
The numerical simulation parameters are presented in Table \ref{tab:SimParam}. 
\begin{table}[t!]
    \centering
    \footnotesize
    \caption{Simulation Parameters}
    \begin{tabular}{l|c|c}
    \toprule
        \textbf{Parameter} &  \textbf{Symbol} & \textbf{Value(s)}\\
        \hline
        Carrier frequency & $f$  & $30$ GHz \\
        Bandwidth & $B$ & $1$ GHz\\
        Transmitted power & $P_{tot}$ & $43$ dBm\\
        Number of subcarriers & $M$ & $1000$\\
        Number of symbols & $N$ & $1000$\\
        Subcarrier Spacing & $\Delta f$ & $1$ MHz \\
        Symbol duration & $T$ & $1$ $\mu$s \\
        OFDM occupancy ratio & $\mu$ & $25,50$ \%\\
        Range & $R$ & $50$ m\\
        Block Size & $N_b$ & 10\\
        SE threshold & $\bar{\eta}$ & 4 bits/s/Hz\\
        \bottomrule
    \end{tabular}
    \label{tab:SimParam}
\end{table}
\begin{figure}[b!]
    \centering
    \includegraphics[width=0.7\columnwidth]{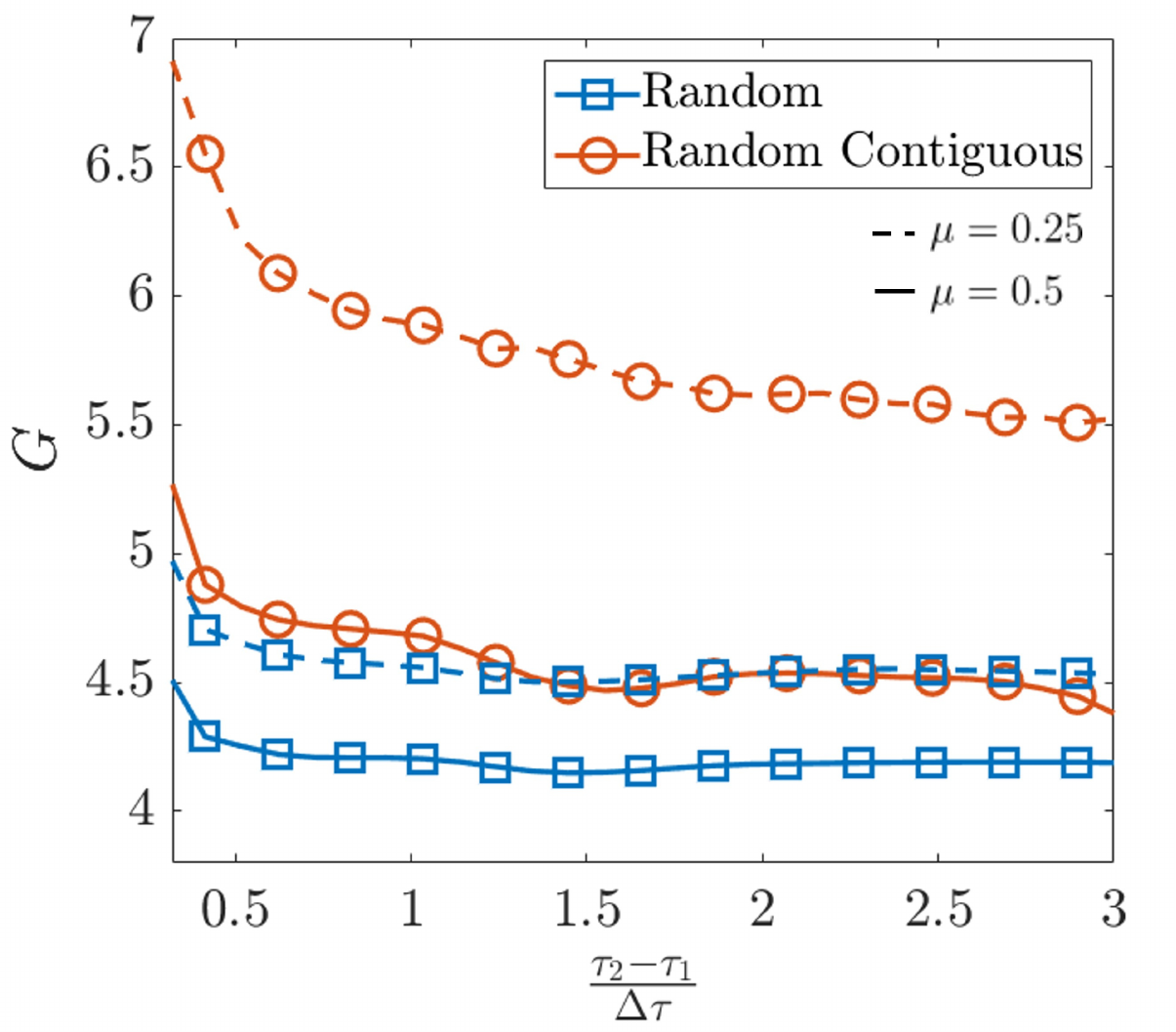}
     \caption{CRB gain of the proposed optimized waveform w.r.t the standard random and random contiguous resource scheduling.}\label{fig:gain}
\end{figure}
By considering constant power allocation $\sigma^2$ per resource and ensuring the bandwidth occupancy $\mu$ for both the proposed approach and the benchmarks, all methods reach the SE threshold $\bar{\eta}$ specified in equation \eqref{eq:prob1_constraint1}.

The first numerical result assesses the CRB gain on delay estimation of two close targets provided by the proposed ISAC waveform design, compared to benchmarks. The CRB gain is defined as:
\begin{align}
    G = \frac{\mathrm{tr}(\mathbf{C}_{\tau, rand})}{\mathrm{tr}(\mathbf{C}_{\tau,opt})} ,
\end{align}
where $\mathbf{C}_{\tau,rand}$ denotes the CRB matrix obtained through the benchmark methods, and $\mathbf{C}_{\tau,opt}$ is the one achievable with the proposed optimized waveform. A gain $G>1$ means a decrease in the CRB for the proposed waveform compared to benchmarks, thus a practical benefit. Assessing the CRB on Doppler estimation provides similar results, thus it is not reported for brevity. Figure \ref{fig:gain} shows the trend of the CRB gain $G$ by varying the inter-delay spacing between the two targets, defined as $|\tau_2 - \tau_1|/\Delta \tau$ (normalized to the delay resolution $\Delta \tau$). With the resource occupancy factor $\mu = 0.25$, the proposed ISAC waveform outperforms the benchmarks by showing a CRB gain $5 \times$ and $7\times$. This underscores the effectiveness of the proposed waveform in discerning closely spaced targets. Additionally, a substantial improvement of around $5 \times$ is observed at $\mu = 0.5$. 

\begin{figure}[t!]
    \centering
    \includegraphics[width=0.85\columnwidth]{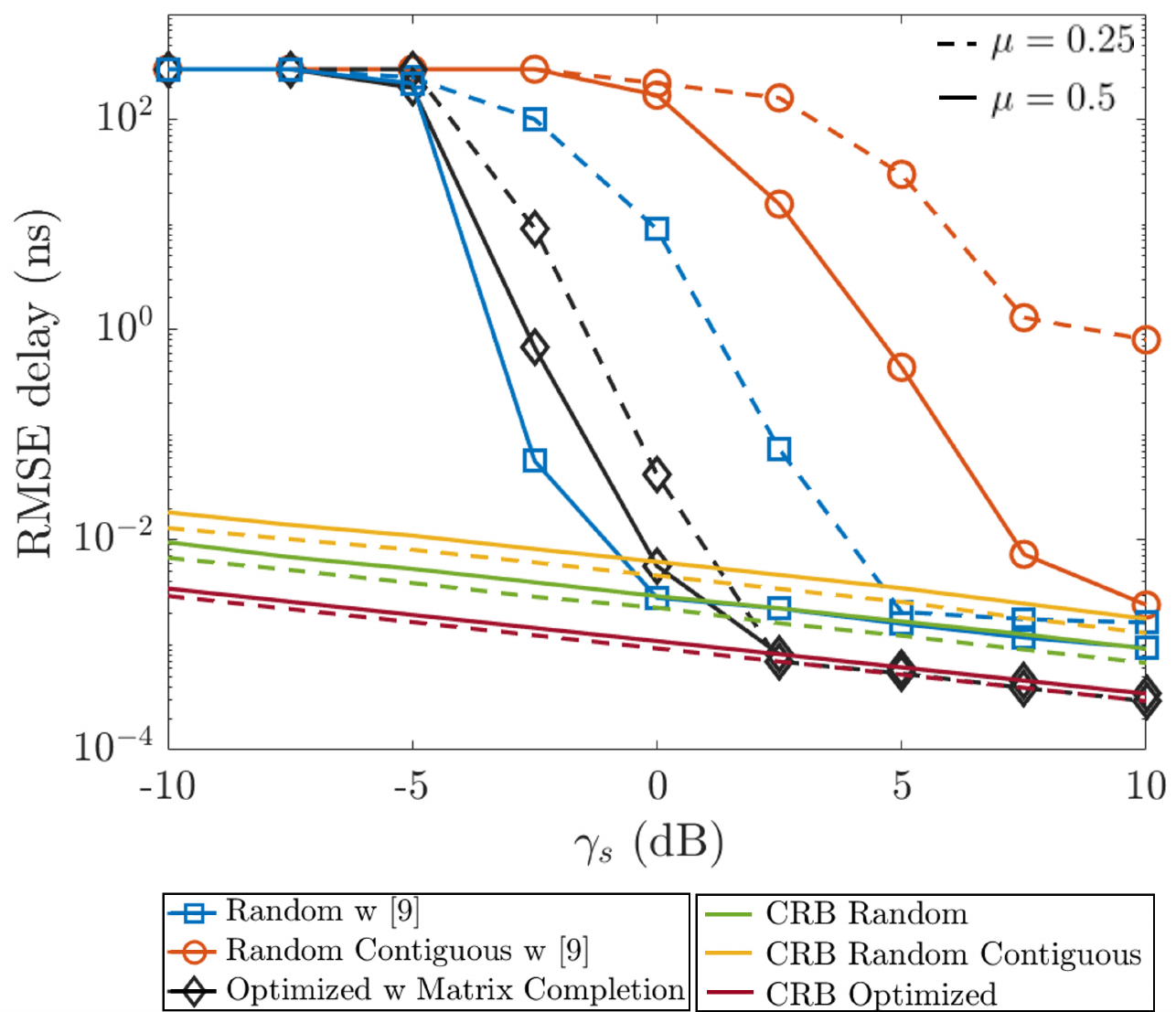}
     \caption{Delay performance estimation of the proposed waveform with respect to the standard random and random contiguous resource allocation with $\mu = 0.25$ and $\mu = 0.5$.}\label{fig:err}
\end{figure}

Figure \ref{fig:err}, shows the root mean square error (RMSE) for the delay estimation achieved by the proposed method and the benchmarks by considering $\mu = 0.25$ and $\mu = 0.5$, a constant total power allocation within the bandwidth ($P_{tot}$ = 43 dBm) and varying the sensing SNR, defined as 
\begin{align}
{\gamma}_{s} = \frac{\sigma^2\,||\boldsymbol{\beta}||_2^2  }{g_s\,\sigma_w^2},
\end{align}
with $\mathbf{\beta} = [\beta_1, \beta_2]$ and $g_s$ representing the processing gain.


 At $\mu = 0.25$, the proposed interpolation method enables to reach the CRB at high ${\gamma}_{s}$, thereby enhancing the performance compared to standard random and random contiguous resource allocation approaches with the linear interpolation in \cite{b7}. The random contiguous scheduling struggles to reach the CRBs; linear interpolation fails in this scenario. Meanwhile, the random waveform, allocating resources per individual resource, exhibits satisfactory performance only at $\mu = 0.5$, ensuring sufficient samples for linear interpolation. Numerical outcomes emphasize the estimation enhancements offered by the proposed waveform under stringent bandwidth occupancy constraints ($\mu = 0.25$), while for $\mu = 0.5$, standard random resource allocation with linear interpolation ensures commendable performance. 

\section{Conclusion}\label{sect:conclusion}

The paper introduces an ISAC OFDM waveform, aiming to reduce delay and Doppler CRBs for nearby dual targets, considering communication efficiency and resource occupancy constraints. Idle resources across time and frequency significantly affect waveform efficacy, leading to suboptimal ambiguity functions with high sidelobes. An interpolation method for the sensing channel, using matrix completion via Schatten $p$-norm approximation, is proposed and it addresses resource occupancy factors below 100\%. The proposed waveform improves CRBs by $7\times$ and $5\times$ compared to standard scheduling methods for closely placed targets. Numerical analysis highlights the enhanced estimation capability under strict bandwidth constraints, while standard methods perform well with higher bandwidth occupancy.

\appendices
\section{CRB for two coupled targets}\label{app:CRB}
We report the CRBs of delay and Doppler estimation in the two-target assumption according to the methodology detailed in~\cite{b5}. We can rewrite \eqref{eq:RxsignalBS} in vector form 
\begin{equation}
    \mathbf{r} = \mathbf{x} \odot  \sum_k \beta_k \odot (\mathbf{d}_{\tau,k}\otimes \mathbf{d}_{\nu,k}) + \mathbf{w},
\end{equation}
where
\begin{align}\label{eq:tau_nu_vec}
    \mathbf{d}_{\tau,k} & = \left[e^{j\pi M \Delta f \tau_k} ,...,1,...,e^{-j\pi (M-1)\Delta f \tau_k}\right]^T\\
    \mathbf{d}_{\nu,k} & = \left[e^{-j\pi N T \nu_{k}} ,...,1,...,e^{j\pi (N-1) T \nu_{k}}\right]^T.
\end{align}
denote delay and Doppler channel responses for $k$-th target and $\mathbf{a}$ = $\mathrm{vec}(\mathbf{A})$.
The Fisher Information Matrix (FIM) $\mathbf{F} \in \mathbb{R}^{2K \times 2K}$ is
\begin{equation}\label{eq:fisher}
 \mathbf{F} = \mathbb{E}_{\mathbf{s}}\{\mathbf{F}(\mathbf{s})\} = \begin{bmatrix}
     \mathbf{F}_{\mathbf{\tau}} \,\mathbf{F}_{\mathbf{\tau}\mathbf{\nu}}\\ \mathbf{F}^{T}_{\mathbf{\tau}\mathbf{\nu}}\,\mathbf{F}_{\mathbf{\nu}} 
 \end{bmatrix}
\end{equation}
where single blocks are reported in \eqref{eq:CRB_tau1tau2}-\eqref{eq:CRB_tau1nu2},
\begin{figure*}
\begin{equation}\label{eq:CRB_tau1tau2}
    [\mathbf{F}_{\tau}]_{k,\ell} = \frac{2 \sigma^2}{{\sigma_w^2}}\, \mathbf{a}^T\Re\left\{4\pi^2 \beta_k\beta_\ell^*\Delta f^2 (\mathbf{m}\hspace{-0.05cm}\odot\hspace{-0.05cm}\mathbf{m}\odot \mathbf{d}_{\tau,k \ell} ) \otimes \mathbf{d}_{\nu,k \ell}\right\} 
\end{equation}
\begin{equation}\label{eq:CRB_nu1nu2}
    [\mathbf{F}_{\nu}]_{k,\ell}= \frac{2 \sigma^2}{\sigma_w^2} \,\mathbf{a}^T \Re\left\{(\mathbf{d}_{\tau,k \ell} \otimes 4\pi^2 T^2 \beta_k\beta_\ell^*(\mathbf{n}\odot\mathbf{n})\odot \mathbf{d}_{\nu,k \ell}) \right\}
\end{equation}
\begin{equation}\label{eq:CRB_tau1nu2}
    [\mathbf{F}_{\nu,\tau}]_{k,\ell}= \frac{2 \sigma^2}{\sigma^2_w} \mathbf{a}^T \mathcal{R}\left\{ 4 \pi^2 \beta_k \beta^*_\ell  \left(\Delta f (\mathbf{m}\odot \mathbf{d}_{\tau,k \ell}) \otimes T (\mathbf{n}\odot \mathbf{d}_{\nu,k \ell})\right)\right\}
\end{equation}
\hrulefill
\end{figure*}
in which
\begin{align}
    \mathbf{n}&=\left[-\frac{N}{2},...,\frac{N}{2}-1\right]^\mathrm{T}\\
    \mathbf{m}&=\left[-\frac{M}{2},...,\frac{M}{2}-1\right]^\mathrm{T}
\end{align}
are the time and frequency index vectors while
\begin{equation}
\begin{split}
    \mathbf{d}_{\tau,k \ell} = \mathrm{diag}(\mathbf{d}_{\tau,k} \mathbf{d}_{\tau, \ell}^H),\,\,\,\,\,
    \mathbf{d}_{\nu,k \ell} = \mathrm{diag}(\mathbf{d}_{\nu,k} \mathbf{d}_{\nu, \ell}^H)
\end{split}
\end{equation}
are the cross-coupled delay and Doppler channel responses.
Under the assumption of $\mathbf{F}_{\nu}$ and $\mathbf{F}_{\tau}$ being non-singular, the CRB for delay and Doppler estimation can be acquired by inverting the FIM as follows:
\begin{align}
    \mathbf{C}_\tau & = (\mathbf{F}_{\tau}-\mathbf{F}_{\nu,\tau}^{\mathrm{T}}\mathbf{F}_{\nu}^{-1} \mathbf{F}_{\nu,\tau})^{-1}\\
    \mathbf{C}_\nu & = (\mathbf{F}_{\nu}-\mathbf{F}_{\nu,\tau}^{\mathrm{T}}\mathbf{F}_{\tau}^{-1} \mathbf{F}_{\nu,\tau})^{-1}.
\end{align}

\section*{Acknowledgment}
This article was supported by the European Union under the Italian National Recovery and Resilience Plan (NRRP) of NextGenerationEU, partnership on “Telecommunications of the Future” (PE00000001 - program “RESTART”).

\end{document}